\documentclass[12pt]{article}

\begin{document}

\title{The one loop measure in the Group Theoretic approach to String Theory: the case of two finite fixed points}

\author{Leonidas Sandoval Junior\\ Department of Mathematics\\ Center for Technological
Sciences\\ UDESC - Universidade do Estado de Santa Catarina\\ Campus Universit\'ario Prof.
Avelino Marcante\\ Joinville - SC - Brazil, CEP 89223-100}

\maketitle

\begin{abstract}
The measure for the one loop scattering of one and $N$ bosonic strings is calculated using the Group Theoretic approach to String Theory. The calculation is done for the case when the
projective transformation associated with the loop can be parametrized in terms of two finite fixed points and the multiplier.
\end{abstract}

\noindent PACS number: 11.25

\noindent Key words: strings, scattering, bosonic, measure, loop, amplitude.

\vskip 1 cm

The Group Theoretic approach to String Theory was developed by A. Neveu and P. C. West
\cite{1}-\cite{5}, where it was used to calculate the scattering amplitude of bosonic strings in all orders of perturbation theory. It has also been used by P. C. West and M. Freeman \cite{6}, \cite{7}, in the study of the ghosts part ({\sl ghost} in the BRST quantization sense) of the scattering of bosonic strings and by A. Neveu, P. C. West and M. Freeman \cite{8} in the study of the scattering amplitudes of superstrings (Neveu-Schwarz and Ramond).

In the Group Theoretic approach, the scattering amplitude $W$ of $N$ physical states $|\chi
_1\rangle $, $|\chi _2\rangle $, $\dots $, $|\chi _N\rangle $ is defined as
\begin{equation}
W\equiv \int \left( \prod_{k=1}^ndz_k\right) \left( \prod_{r=1}^{3(g-1)}dv_r\right) \bar
f(z_i,v_r)V(z_i,v_r)|\chi _1\rangle \dots |\chi _N\rangle \ ,
\end{equation}
where $\bar f(z_i,v_r)$ is the integration measure, or {\sl measure}, for short, and
$V(z_i,v_r)$ is the {\sl vertex}, which can be represent in the oscillator basis as
\begin{equation}
V(z_i,v_r)=\left( \prod_{i=1}^N{ }_i\! \langle 0|\right) h(\alpha ^{\mu i}_m,z_i,v_r)\ ,
\end{equation}
where $\alpha ^{\mu i }_m$ ($m>0$) are annihilation operators of scalar fields acting on $N$
vacuum states $|0\rangle _i$. The moduli of the Riemann surface with $g$ loops and $N$
attached legs (i.e. strings) are the Koba-Nielsen variables $z_i$ ($i=1,\dots ,N$), one for
each leg, and $v_r$ ($r=1,\dots ,3(g-1)$), three for each loop, minus 3 that are fixed in
order to gauge-fix the modular invariance of the theory.

By modular invariance we mean the invariance under transformations
\begin{equation}
P_n(z)=\frac{a_nz+b_n}{c_nz+d_n}
\end{equation}
where $n=1,\dots ,g$ and $a_nd_n-b_nc_n=1$ for every $n$. In the case of just one loop we have just one of these projective transformations, which can be written just as
\begin{equation}
P(z)=\frac{az+b}{cz+d}
\end{equation}
where $ad-bc=1$. This transformation can be parametrized in terms of three other variables
$\alpha $, $\beta $ and $w$. The variables $\alpha $ and $\beta $ are called {\sl fixed
points} since $P(\alpha )=\alpha $ and $P(\beta )=\beta $. These are given by
\begin{equation}
\alpha =\frac{b}{d-a}\ \ ,\ \ \beta =\infty \ \ {\rm when}\ \  c=0 \ \ {\rm and}\ \  a\neq d
\end{equation}
or
\begin{equation}
\alpha =\frac{a-d+\sqrt{\Delta }}{2c}\ \ ,\ \ \beta =\frac{a-d-\sqrt{\Delta }}{2c}
\end{equation}
where
\begin{equation}
\Delta =(a+d)^2-4\ ,\ \Delta \neq 0 \ \ {\rm and}\ \  c\neq 0\ .
\end{equation}
The third variable is called the {\sl multiplier} and is defined by
\begin{equation}
w=\frac{a}{d}\ \ ,\ \ c=0\ \ ,\ \ a\neq d
\end{equation}
or
\begin{equation}
w=\frac{a-c\alpha }{a-c\beta }\ \ ,\ \ c\neq 0\ \ ,\ \ \Delta \neq 0\ .
\end{equation}
In terms of these variables, the projective transformation $P(z)$ is written like
\begin{eqnarray}
 & & P(z)=w(z-\alpha )+\alpha \ \ ,\ \ c=0\ \ ,\ \ a\neq d\ ,\\
 & & P(z)=\frac{\alpha (z-\beta )-w\beta (z-\alpha )}{(z-\beta )-w(z-\alpha )}\ \ ,\ \ c\neq
0\ .
\end{eqnarray}

In the Group Theoretic approach, the vertex and the measure can be calculated by imposing
certain conditions on them \cite{5}. In this article we shall be concerned with the
calculation of the measure for one loop scattering amplitudes. This has already been done for the case where $c=0$ and $a\neq d$, i.e. the case when we have one finite fixed point $\alpha $ and the other ($\beta $) set to infinity. This works well for the one loop case, but cannot be used for the general multiloop case, where we have to use pairs of finite fixed points for each loop. It becomes interesting then to calculate the measure for the scattering amplitude in one loop for the case of two finite fixed points. This is what is done article here.

In the Group Theoretic approach, the measure can be calculated by using some functions that
change the moduli $z_i$ and $v_r$ of the Riemann surface in order to obtain identities
involving the generators of the Virasoro algebra. These relations are then used with the fact that the scattering amplitudes of zero norm states are zero in order to obtain some
differential equations for the measure. This calculation has been done \cite{2} for the case
when we choose to set one of the fixed points at infinity and the other fixed point is set to have a finite constant value. One of the Koba-Nielsen variables $z_i$ is also fixed so that the modular invariance of the theory is gauged. We then have functions that transform the remaining Koba-Nielsen variables infinitesimally and other functions that transform the
multiplier $w$ infinitesimally. Since we have fixed the values of both fixed points, the
resulting measure will not depend on them. The result obtained for the one loop measure was
\begin{equation}
\bar f=w^{-2}\left[ \prod_{n=1}^\infty (1-w^n)\right] ^{-24}\ .
\end{equation}
This result is independent of the number of legs.

When we calculate the scattering of $N$ bosonic strings, we can also consider the case when
instead of fixing one of the Koba-Nielsen variables and the two finite fixed points, we want
to fix three of the Koba-Nielsen variables and leave the fixed points undetermined. In this
case, besides considering the moduli-changing functions that change the remaining $N-3$
Koba-Nielsen variables infinitesimally and the functions that change the multiplier $w$, we
must also take into account the functions that change the two fixed points.

In this case, the functions that change the $N-3$ unfixed Koba-Nielsen variables by $\delta z_i$ are given by
\begin{eqnarray}
 & & f_i^{(i)}(\xi _i)=\frac{\delta z_i}{\epsilon }\frac{(\xi _i-\alpha )(\xi _i-\beta )}{\alpha \beta }\ ,\\
 & & f_j^{(i)}(\xi _j)=0\ ,\ j\neq i\ ,
\end{eqnarray}
for some $i=1,2,\dots ,N-3$ (we are choosing the Koba-Nielsen variables that are being fixed
to be the last three ones, i.e. $N-2$, $N-1$ and $N$), where $\epsilon $ is an infinitesimal constant. In the expression above, $\xi _i$ are functions such that $\xi _i(z_i)=0$ (the simplest case is $\xi _i=z-z_i$). If we substitute these functions into the relation
\begin{equation}
V\left[ \sum_{j=1}^N\int d\xi _j\ \epsilon f_j^{(i)}T^j(\xi _j)+c_1^{(i)}\right] =\delta
z_i\frac{\partial V}{\partial z_i}\ ,\ i=1,\dots ,N-3\ ,
\end{equation}
we obtain, after verifying that $c^{(i)}=0$, what can be done by acting with the expression above on a vacuum state,
\begin{equation}
\label{eq16}
VL_{-1}^i=V\left( \frac{\alpha +\beta }{\alpha \beta }L_0^i-\frac{1}{\alpha \beta
}L_1^i\right) +\frac{\partial V}{\partial z_i}\ ,\ i=1,\dots ,N-3\ .
\end{equation}
This relation shall be used later.

The functions that change the multiplier by $\delta w$ are given by
\begin{eqnarray}
 & & f_{wi}^{(i)}(\xi _i)=\frac{\delta w}{\epsilon w}\frac{(\xi _i-\alpha )(\xi _i-\beta )}{\alpha -\beta }\bar \zeta (v_i)\ ,\\ 
 & & f_{wj}^{(i)}(\xi _j)=\frac{\delta w}{\epsilon w}\frac{(\xi _j-\alpha )(\xi _j-\beta )}{\alpha -\beta }\bar \zeta (v_j+\zeta _j-\zeta _i)\ ,
\end{eqnarray}
where $v_i=\ln \left( \frac{\xi _i-\alpha }{\xi _i-\beta }\right) -\ln \left( \frac{\alpha }{\beta }\right) $ are the first abelian integrals. This function will generate the same dependence of the measure on the variable $w$ as in the case considered before. The only difference is that now the measure may also depend on the fixed points. So, we write the measure, for now, like
\begin{equation}
\label{eq19}
\bar f=w^{-2}\left[ \prod_{n=1}^\infty (1-w^n)\right] ^{-24}g(\alpha ,\beta )
\end{equation}
where $g(\alpha ,\beta )$ is a function of the two fixed points.

In order to determine the dependence of the measure on the fixed points, we must consider the functions that change these moduli. The functions that change the fixed point $\alpha $ by an infinitesimal $\delta \alpha $ must satisfy the identities
\begin{eqnarray}
 & & f_{\alpha i}^{(j)}(P^j\xi _j)=P'^j(\xi _j)\left[ f_{\alpha j}^{(i)}(\xi _j)-\frac{(1-w)}{w(\alpha -\beta )^2}(\xi _j-\beta )^2\frac{\delta \alpha }{\epsilon }\right] \ ,\\ 
 & & f_{\alpha j}^{(i)}({\rm e}^{2\pi i}\xi _j)=f_{\alpha j}^{(i)}(\xi _j)\ ;
\end{eqnarray}
and the functions that change the fixed point $\beta $ by an infinitesimal $\delta \beta $ must satisfy
\begin{eqnarray}
 & & f_{\beta i}^{(j)}(P^j\xi _j)=P'^j(\xi _j)\left[ f_{\beta j}^{(i)}(\xi _j)-\frac{(1-w)}{(\alpha -\beta )^2}(\xi _j-\alpha )^2\frac{\delta \beta }{\epsilon }\right] \ ,\\ 
 & & f_{\beta j}^{(i)}({\rm e}^{2\pi i}\xi _j)=f_{\beta j}^{(i)}(\xi _j)\ .
\end{eqnarray}
The functions that satisfy these relations are, respectively,
\begin{equation}
f_{\alpha j}^{(i)}(\xi _j)=\frac{\delta \alpha }{\epsilon }\frac{1}{(\alpha -\beta )^2}\left[ c_\alpha (\xi _j-\alpha )(\xi _j-\beta )-(\xi _j-\beta )^2\right] \ ,
\end{equation}
and
\begin{equation}
f_{\beta j}^{(i)}(\xi _j)=\frac{\delta \beta }{\epsilon }\frac{1}{(\alpha -\beta )^2}\left[ c_\alpha (\xi _j-\alpha )(\xi _j-\beta )-(\xi _j-\alpha )^2\right] \ ,
\end{equation}
where $c_\alpha $ and $c_\beta $ are arbitrary constants such that $c_\alpha \neq \frac{\beta }{\alpha }$ and $c_\beta \neq \frac{\beta }{\alpha }$.

We now substitute these functions into the identities \cite{3}
\begin{eqnarray}
 & & V\left[ \sum_{j=1}^N\int d\xi _j\ \epsilon f_{\alpha j}^{(i)}(\xi _j)T^j(\xi _j)+c_3\right] =\delta \alpha \frac{\partial V}{\partial \alpha }\ ,\\ 
 & & V\left[ \sum_{j=1}^N\int d\xi _j\ \epsilon f_{\beta j}^{(i)}(\xi _j)T^j(\xi _j)+c_4\right] =\delta \beta \frac{\partial V}{\partial \beta }\ ,
\end{eqnarray}
where $c_3$ and $c_4$ are constants that can be calculated to be zero by acting with the expressions above on a vacuum state. What we obtain after integration is
\begin{eqnarray}
 & & V\sum_{j=1}^N\left\{ \beta (c_\alpha \alpha -\beta )L_{-1}^j-\left[ c_\alpha (\alpha +\beta )-2\beta \right] L_0^j+(c_\alpha -1)L_1^j\right\} \nonumber \\ 
\label{eq28}
 & & =(\alpha -\beta )^2\frac{\partial V}{\partial \alpha }\ ,\\ 
 & & V\sum_{j=1}^N\left\{ \alpha (c_\beta \beta -\alpha )L_{-1}^j-\left[ c_\beta (\alpha +\beta )-2\alpha \right] L_0^j+(c_\beta -1)L_1^j\right\} \nonumber \\ 
\label{eq29}
 & & =(\alpha -\beta )^2\frac{\partial V}{\partial \beta }\ .
\end{eqnarray}

From expression (\ref{eq28}) we have
\begin{eqnarray}
 & & VL_{-1}^i=V\left[ \frac{c_\alpha (\alpha +\beta )-2\beta }{\beta (c_\alpha \alpha -\beta )}L_0^i-\frac{c_\alpha -1}{\beta (c_\alpha \alpha \beta )}L_1^j\right] \nonumber \\ 
 & & -V\sum_{\scriptstyle j=1\atop \scriptstyle j\neq i}\left[ L_{-1}^j-\frac{c_\alpha (\alpha +\beta )-2\beta }{\beta (c_\alpha \alpha -\beta )}L_0^i+\frac{c_\alpha -1}{\beta (c_\alpha \alpha -\beta )}L_1^j\right] \nonumber \\ 
 & & +\frac{(\alpha -\beta )^2}{\beta (c_\alpha \alpha -\beta )}\frac{\partial V}{\partial \alpha }\ .
\end{eqnarray}
Using identity (\ref{eq16}) for all $j\neq i$ we then obtain the expression
\begin{eqnarray}
 & & VL_{-1}^i=V\left[ \frac{c_\alpha (\alpha +\beta )-2\beta }{\beta (c_\alpha \alpha -\beta )}L_0^i-\frac{c_\alpha -1}{\beta (c_\alpha \alpha \beta )}L_1^j\right] \nonumber \\ 
 & & -V\sum_{\scriptstyle j=1\atop \scriptstyle j\neq i}\left[ \frac{\alpha -\beta }{\alpha (c_\alpha \alpha -\beta )}L_0^i-\frac{\alpha -\beta }{\alpha \beta (c_\alpha \alpha -\beta )}L_1^j\right] \nonumber \\ 
 & & -V\sum_{\scriptstyle j=1\atop \scriptstyle j\neq i}\frac{\partial V}{\partial z_j} +\frac{(\alpha -\beta )^2}{\beta (c_\alpha \alpha -\beta )}\frac{\partial V}{\partial \alpha }\ .
\end{eqnarray}
If we act with the identity above on a state $|k\rangle _{1i}\dots |\Psi \rangle _N$, where $|k\rangle _{1i}$ is a state with norm zero, we then have
\begin{eqnarray}
 & & VL_{-1}^i|k\rangle _{1i}\dots |\Psi \rangle _N=\left[ V\frac{(N-1)(\alpha -\beta )}{\alpha (c_\alpha \alpha -\beta )}\right. \nonumber \\ 
\label{eq32}
 & & \left. -V\sum_{\scriptstyle j=1\atop \scriptstyle j\neq i}\frac{\partial V}{\partial z_j} +\frac{(\alpha -\beta )^2}{\beta (c_\alpha \alpha -\beta )}\frac{\partial V}{\partial \alpha }\right] |k\rangle _{1i}\dots |\Psi \rangle _N\ .
\end{eqnarray}

Let us now consider the following expression:
\begin{equation}
\int \left( \prod_{k=1}^{N-3}dz_k\right) dwd\alpha d\beta \ \bar fVL_{-1}^i|k\rangle _{1i}\dots |\Psi \rangle _N=0\ ,
\end{equation}
which comes from the fact that the scattering amplitude of states with zero norm should be zero. Using expression (\ref{eq32}) in the expression above, we obtain
\begin{eqnarray}
 & & \int \left( \prod_{k=1}^{N-3}dz_k\right) dwd\alpha d\beta \ \bar f\left[ V\frac{(N-1)(\alpha -\beta )}{\alpha (c_\alpha \alpha -\beta )}\right. \nonumber \\ 
 & & \left. -\sum_{\scriptstyle j=1\atop \scriptstyle j\neq i}\frac{\partial V}{\partial z_j} +\frac{(\alpha -\beta )^2}{\beta (c_\alpha \alpha -\beta )}\frac{\partial V}{\partial \alpha }\right] |k\rangle _{1i}\dots |\Psi \rangle _N=0\ .
\end{eqnarray}
After integrating by parts on the Koba-Nielsen variables (using the fact that the measure is independent of them), this simplifies to
\begin{equation}
\int dwd\alpha d\beta \ \bar f\left[ V\frac{(N-1)(\alpha -\beta )}{\alpha (c_\alpha \alpha -\beta )}+\frac{(\alpha -\beta )^2}{\beta (c_\alpha \alpha -\beta )}\frac{\partial V}{\partial \alpha }\right] |k\rangle _{1i}\dots |\Psi \rangle _N=0\ .
\end{equation}
Integrating by parts over $\alpha $, we then have
\begin{equation}
\int dwd\alpha d\beta \left\{ \frac{\partial }{\partial \alpha }\left[ \frac{(\alpha -\beta )^2}{\beta (c_\alpha \alpha -\beta )}\bar f\right] -\frac{(N-1)(\alpha -\beta )}{\alpha (c_\alpha \alpha -\beta )}\bar f\right\} V|k\rangle _{1i}\dots |\Psi \rangle _N=0\ ,
\end{equation}
where we take
\begin{equation}
\label{eq37}
\left. \frac{(\alpha -\beta )^2}{\beta (c_\alpha \alpha -\beta )}\right| _{\alpha =-\infty }^\infty =0\ .
\end{equation}
This implies the following differential equation for $\bar f$:
\begin{equation}
\frac{\partial }{\partial \alpha }\left[ \frac{(\alpha -\beta )^2}{\beta (c_\alpha \alpha -\beta )}\bar f\right] =\frac{(N-1)(\alpha -\beta )}{\alpha (c_\alpha \alpha -\beta )}\bar f\ ,
\end{equation}
which can be written as
\begin{equation}
\frac{\partial }{\partial \alpha }\ln \left[ \frac{(\alpha -\beta )^2}{\beta (c_\alpha \alpha -\beta )}\bar f\right] =\frac{\beta (N-1)}{\alpha (\alpha -\beta )}\ .
\end{equation}
Integrating this differential equation we then obtain
\begin{equation}
\ln \left[ \frac{(\alpha -\beta )^2}{\beta (c_\alpha \alpha -\beta )}\bar f\right] =(N-1)\ln \left| \frac{\alpha -\beta }{\alpha }\right| +f(w,\beta ) \ ,
\end{equation}
or
\begin{equation}
\bar f=\frac{c_\alpha \alpha -\beta }{(\alpha -\beta )^2}\left| \frac{\alpha -\beta }{\alpha }\right| ^{N-1}g(w,\beta ) \ ,
\end{equation}
where $f(w,\beta )$ is a function of the other two moduli and $g(w,\beta )={\rm e}^{f(w,\beta )}$. This measure satisfies condition (\ref{eq37}).

The determination of the dependence of the measure on the fixed point $\beta $ goes the same way. At the end, what we obtain is
\begin{equation}
\bar f=\frac{c_\beta \beta -\alpha }{(\alpha -\beta )^2}\left| \beta (\alpha -\beta )\right| ^{-(N-1)}h(w,\alpha ) \ ,
\end{equation}
where $h(w,\alpha )$ is a function that must be compatible with the results obtained before. Compatibility with the expression for the measure (\ref{eq19}) obtained from the equations for $w$ implies that we must have
\begin{eqnarray}
\label{eq43}
 & & \bar f=\frac{c_\alpha \alpha -\beta }{(\alpha -\beta )^2}\left| \frac{\alpha -\beta }{\alpha }\right| ^{N-1}w^{-2}\left[ \prod_{n=1}^\infty (1-w^n)\right] ^{-24}f(\beta ) \ ,\\ \label{eq44}
 & & \bar f=\frac{c_\beta \beta -\alpha }{(\alpha -\beta )^2}\left| \beta (\alpha -\beta )\right| ^{-(N-1)}w^{-2}\left[ \prod_{n=1}^\infty (1-w^n)\right] ^{-24}g(\alpha ) \ ,
\end{eqnarray}
where $f(\beta )$ and $g(\alpha )$ must make both expressions compatible. 

The only way to make the two expressions for the measure compatible is by choosing suitable values for the constants $c_\alpha $ and $c_\beta $. This freedom of choice can also be used in order to eliminate the dependence of the measure on the two fixed points. This would make the expression for the case of two finite fixed points identical to the one previously obtained for the case of one finite fixed point and the other one fixed at infinity. Choosing
\begin{eqnarray}
 & & c_\alpha =\frac{1}{\alpha }\left( \beta +\left| \frac{\alpha }{\alpha -\beta }\right| ^{N-1}\right) \ ,\\ 
 & & c_\beta =\frac{1}{\beta }\left[ \alpha +\left| \frac{1}{\beta (\alpha -\beta )}\right| ^{N-1}\right] \ ,
\end{eqnarray}
has the effect of turning expressions (\ref{eq43}) and (\ref{eq44}) into
\begin{equation}
\bar f=w^{-2}\left[ \prod_{n=1}^\infty (1-w^n)\right] ^{-24}\ .
\end{equation}

As can be seen from the results obtained, suitable choices for the constants $c_\alpha $ and $c_\beta $ can make the measure for the case of two finite fixed points identical to the one with one finite fixed point and the other set to infinity. Other choices are also possible, but are restricted by compatibility of the expressions for the measure obtained from the equations for the two finite fixed points. All expressions obtained this way are independent of the number of legs, as should be expected.

\vskip 0.5 cm

\noindent {\large \bf Acknowledgements}

\vskip 0.3 cm

The author would like to thank Professor P. C. West, under whose supervision this work was completed, and to CAPES and CNPq (Brazilian government) for financing this work. G.E.D.!

\end{document}